\documentstyle[12pt]{article}

\author{L.S.F. Olavo\\
Departamento de Fisica ,Universidade de Brasilia - UnB\\
70910-900, Brasilia - D.F. - Brazil}
\title{Quantum Mechanics as a Classical Theory\\
XIII: The Tunnel Effect
}
\date{September 3, 1996}

\begin{document}

\maketitle
\begin{abstract}
In this continuation paper we will address the problem of tunneling. We will
show how to settle this phenomenon within our classical interpretation. It
will be shown that, rigorously speaking, there is no tunnel effect at all.
\end{abstract}

\section{Introduction}

Since the very moment when one is introduced to the quantum formalism {\it %
via} Schr\"odinger equation, he takes contact, as an initiation ritual, with
the problems involving those potentials considered simple---those which
mathematical solutions is sufficiently feasible, furnishing however some
acceptable physical content.

The student is then first exposed to problems related with the ''square
potentials'' (step, barrier, etc.) and only later will face problems like
the harmonic oscillator, the hydrogen atom, etc.

It is with the square potentials that he first is put into contact with the
intriguing tunneling effects. It is also with these first ''naive''
applications that it becomes established the fundamental distinction between
the theoretical quantum level (where such tunneling effects are expected)
and the classical one (where such effects cannot, supposedly, be expected).
This first mental exercise, by its very simplicity, has such a powerful
effect over the conceptions of those involved with it that the impressions,
left behind by the interpretation usually superimposed on it by the accepted
epistemology, remain in the spirit of the initiate for all his future
carrier.

This is precisely the motivation for these first steps: the indelible
fixation of a determinate world scheme in the spirit of the future
generation of scientists---the forge of a consensus. Such exercises, hence,
are exemplary and, hardly, ''naive''.

In the present series of papers \cite{1}-\cite{12} we are trying to show
that it is possible to superimpose to the formal apparatus of quantum
mechanics a completely classical interpretation, without appealing to anyone
of the usual interpretation aspects of the quantum epistemology (e.g.
observers, wave-particle duality, etc.).

It is the intention of this paper to elucidate the known problem of the
tunnel effect and show that it may also be settled within a purely classical
interpretation. It is unnecessary to say that no appeal to the duality
aspect of matter or any other scheme not founded on the classical
interpretation will be here allowed. We will have to deal only with {\it %
ensembles} of classical particles.

To attain this goal, in the second section we will proceed with a quick
analysis of the mathematical developments found in the literature. We are
interested here with a critical analysis of the square potentials.

In the third section we will present, in a general fashion, the harmonic
oscillator problem in the way it is described in the literature. This
problem addresses the one of tunneling for it presents, following the usual
interpretation, a non-zero probability of finding particles in regions
beyond the classical turning points.

In the forth section the same problem will be dealt with following the
formalism already developed in a number of previous papers. We will then
reveal, with an analysis of the problem on the phase-space, which
interpretation is more adequate to this phenomenon. The classical character
of this phenomenon will be totally established. It is possible that, after
this exposition, those more sensitive, generally inclined to purism, will
find it hard to use the denomination ''tunneling'' to this effect. The
vanishing of the nomenclature, however, will reveal the very weakness of the
interpretation responsible for its significance.

\section{Square Potentials}

The first potentials to which we are first presented when initiating the
study of quantum theory are, for reasons of mathematical simplicity, those
we may classify as square potentials, after their mathematical expression%
\cite{13,14,15}

\begin{equation}
\label{1}V(x)=\left\{ 
\begin{array}{l}
V0
\mbox{ if }b\le x\le c, \\ 0\mbox{ else }
\end{array}
\right. .
\end{equation}

The solutions related with the introduction of such potentials are given by 
\begin{equation}
\label{2}\psi (x)=Ae^{ikx}+Be^{-ikx},
\end{equation}
being $k$ , eventually, complex (e.g. inside the barrier) and where $A$ and $%
B$ are constants.

We may at this point present two sources of problems: the first one is
related with the expression for the potentials while the second is related
with the aspect of the functions which are solutions of the Schr\"odinger
equation resulting from the consideration of these potentials.

Clearly, a potential as the one described in (\ref{1}) does not pertain to
the scope of a quantum formalism as here understood. Indeed, we shall
remember that the Schr\"odinger equation was {\it derived} from the
Liouville equation using the supposition

\begin{equation}
\label{3}\frac{\partial V(x)}{\partial x}\delta x=V(x+\delta x/2)-V(x-\delta
x/2), 
\end{equation}
where $\delta x$ is an infinitesimal displacement. This supposition requires
that the potential has a continuous first order derivative over all its
domain of definition.

This constraint is also related with the very use of the potential inside
the Liouville equation, where it appears inside the derivative sign.

Since these potentials do not satisfy the {\it conditio sine qua non} of the
quantum formalism, they cannot be considered as adequate to represent
situations manageable within the theory. Hence, far from being exemplary for
a first study of the formalism, they are, indeed, by no means,
representative of the kind of phenomena the theory is apt to reveal.

At this stage, some attentive reader might ask how it is possible to say
such a thing if we see at every moment the experiments confirming the
behavior raised by the form of these potentials? If, even unexpected from
realistic situations, it wouldn't be possible to find various potentials
that approximate the square ones as close as one wants?

To answer these two questions it is necessary to go somewhat further into
our discussion. Hence, we will postpone the solution of this apparent
dilemma to the fourth section.

With respect to the solutions we get when we introduce the square potentials
into the Schr\"odinger equation, we may say the following: when the value of 
$k$ in expression (\ref{2}) is real inside an infinite domain of the
variable $x$, it is not possible to find a normalization for these
functions---since they are not $L^2$. Indeed, not being $L^2$ these
functions are not even members of the set of acceptable functions of the
quantum formalism.

Under these circumstances, what we usually do is to appeal to a mathematical
process called ''box normalization'' where we consider the particle as being
in the interior of a cubic box of side $L$, and being normalized within this
finite volume. We make the box volume tend to infinity keeping constant the
overall probability density related with the functions. This is the
mathematical way by means of which we reintroduce the functions (\ref{2}) in
the formal apparatus of quantum formalism.

Even if we decide to accept this mathematical process as licit, it is
possible to ask if the use of such functions may be accommodated within the
interpretation usually superimposed to the quantum formalism.

Indeed, being functions whose value of the energy is completely fixed (e.g. $%
E$), these functions, due to the Heisenberg dispersion relations, loose all
its temporal localization, or saying differently, it is not possible to fix
in any way, coherent with the principles of quantum mechanics, an initial
time for the phenomenon.

Due to all we have said above, we have to absolutely refuse the objectivity
of the phenomena described by the square potentials and their respective
solutions.

It is, however, possible to argue that there exists other examples where the
phenomenon of tunneling appears. One clear such example is the harmonic
oscillator where, we may show, there exists a finite probability of finding
particles in the classical forbidden regions (beyond the classical turning
points), as explained by the usual interpretation.

In the following section we will present in a very brief way the formalism
and interpretation usually associated with this problem.

\section{Harmonic Oscillator}

Within this application the potential is given by

\begin{equation}
\label{5}V(x)=C^2x^2/2,
\end{equation}
where $C$ is a constant. Clearly, this is an acceptable potential for the
quantum formalism and our first complain of the last section is no longer
valid.

The Schr\"odinger equation associated with this potential may be written as

\begin{equation}
\label{6}\frac{d^2\psi (x)}{dx^2}+(\beta -\alpha ^2)\psi (x)=0,
\end{equation}
where 
\begin{equation}
\label{7}\alpha \equiv 2\pi m\nu /\hbar \mbox{ and }\beta \equiv 2mE/\hbar ,
\end{equation}
with $m$ the particle mass, $E$ its energy, $\hbar $ Planck's constant and
the frequency $\nu $ given by 
\begin{equation}
\label{8}\nu ^2=\frac{C^2}{4\pi ^2m}.
\end{equation}

The normalized solutions for this problem are amply known and may be written
as 
\begin{equation}
\label{9}\psi (x)=\left( \frac{\sqrt{\alpha }}{\sqrt{\pi }2^nn!}\right)
^{1/2}e^{-\alpha x^2/2)}H_n(\sqrt{\alpha }x),
\end{equation}
where $H_n(x)$ are the Hermite polynomials. The energies become 
\begin{equation}
\label{10}E_n=(n+1/2)h\nu ,\mbox{   }n=0,1,2,3,...
\end{equation}
where $h=2\pi \hbar $.

When looking at anyone of these solutions we may perceive that it exists a
finite probability of finding particles in the regions beyond the classical
turning points.

We also note that the functions are $L^2$ and hence are not subjected to our
previous criticism, made in the last section.

It then seems to be an obvious conclusion the impossibility of presenting a
classical interpretation to explain this phenomenon. Indeed, in the regions
beyond the classical turning points, the kinetic energy has to be negative
implying a complex momentum.

In the next section we will show how such an interpretation may be give.

\section{Classical Tunnelling}

For the same potential of the previous section we may write the hamiltonian
function as 
\begin{equation}
\label{11}H=\frac{p^2}{2m}+\frac 12C^2x^2.
\end{equation}
If $H_0$ is the initial energy of some particle when submitted to the
potential (\ref{5}), the classical turning points may be written as 
\begin{equation}
\label{12}x_{ret}=\pm \sqrt{2H_0}/C=\pm p_0/\sqrt{M}C,
\end{equation}
where $p_0$ is its initial momentum. These points are obtained making the
final momentum of the particle $p_f$=0.

Taking, without loss of generality, the fundamental state $n=0$ in the
quantum solutions obtained in the last section, we may write 
\begin{equation}
\label{13}H_0=\hbar C/\sqrt{M}\mbox{ and }\psi _0(x)=\pi ^{-1/4}\alpha
^{1/4}e^{-\alpha x^2/2}.
\end{equation}
The probability density in configuration space becomes 
\begin{equation}
\label{14}\rho (x)=(\alpha /\pi )^{1/2}e^{-\alpha x^2}.
\end{equation}
Such probability density may be used to find the probability of having
particles beyond the classical turning points. This probability may be
written as 
\begin{equation}
\label{15}pr^Q=\int_{-\infty }^{-x_{ret}}{\rho (x)}dx+\int_{+x_{ret}}^\infty 
{\rho (x)}dx,
\end{equation}
which gives, performing the integrals, 
\begin{equation}
\label{16}pr^Q=1-erf(\alpha ^{1/2}x_{ret}),
\end{equation}
where $erf(x)$ stays for the error function.

Using now the infinitesimal Wigner-Moyal transformation, as defined in our
previous papers\cite{1}-\cite{12}, we may get the normalized classical
probability distribution function, defined upon phase space, as

\begin{equation}
\label{18}F_0(x,p;t)=\frac 1{\pi \hbar }e^{-ax^2-p^2/a\hbar ^2}. 
\end{equation}

One has to note, however, that this classical distribution implies that
there exists a dispersion in the momentum $\Delta p$ ---and also in the
energy (as we have already pointed out\cite{9}). In this case, within the
interpretation scheme proposed by this series of papers, we have to accept
that there exists particles in the {\it ensemble} with various values of
momenta%
\footnote{Note that it is not essential to go to the classical distribution. The 
quantum distribution in momentum space itself may be used. The calculations with this 
function will give, however, exactly the same results, since this distribution is
equivalent to the lateral distribution in momentum space derived from the classical
function, as was showed in the appendix A of our first paper[1].}.

We may then ask what is the probability of finding particles within this 
{\it ensemble} with initial momentum $p_i$ greater then $p_0$. Such a
calculation is performed by means of the integral 
\begin{equation}
\label{19}pr^{cl}=2\int_{p_0}^\infty {F_0(p)}dp, 
\end{equation}
where 
\begin{equation}
\label{20}F_0(p)=\int_{-\infty }^\infty {F_0(x,p;t)}dx. 
\end{equation}
It is easy to show that the result of this integration gives 
\begin{equation}
\label{21}pr^{cl}=1-erf(\frac{p_0}{\sqrt{\alpha }\hbar }). 
\end{equation}

To have the two results coinciding we must have 
\begin{equation}
\label{22}\frac{p_0}{\sqrt{\alpha }\hbar }=\sqrt{\alpha }x_{ret}.
\end{equation}
Noting that 
\begin{equation}
\label{23}\alpha \hbar =C\sqrt{m},
\end{equation}
we get 
\begin{equation}
\label{24}x_{ret}=\frac{p_0}{\sqrt{m}C},
\end{equation}
which is precisely the expression (classical) to the classical turning point
obtained in (\ref{12}).

This result signifies that, inside the original {\it ensemble}, whose mean
kinetic energy whose energy was $H_0$, there were a number of particles
whose momentum module was greater or equal to the momentum $p_0$, necessary
to go beyond the classical turning points (derived from this{\it \ mean
energy}). Hence, nothing more obvious (and classical) than having these
particles going beyond those classical turning points, since their initial
energy is greater than the energy used to derive these points.

The phenomenon may be completely explained from the classical statistical
point of view, without appealing to the negative kinetic energy concept. or
any other pathology.

We may now answer the questions made in the second section with respect to
the adequacy of using square potentials to simulate real potentials. We have
to consider carefully the fact that some potential, not being of the square
type, no matter how close it is of this format, will never allow solutions
whose energy is free from dispersion. In this case, it will always be
possible to find in the {\it ensemble} a certain number of particles with
high values of the initial momentum $p_i$, even in a very small number (i.e.
with small probability). These are the particles that will surmount the
potential barrier. The solutions of the Schr\"odinger equation will be
absolutely acceptable.

The choice of square potentials simplifies too much the problem denying us
to identify all the subtleties related to it and inducing us to interpret
its formal results in a mistaken manner. Their use in the scope of quantum
theory is then subjected to great criticism. One way to overcome partially
the problem is to use superpositions of plane waves which allow us to
simulate an initial distribution function that presents the desired
deviations.

\section{Conclusions}

In this brief paper we have shown how one may explain, inside the formalism
of quantum mechanics and within the classical statistical {\it ensemble}
interpretation, the widely known phenomenon of tunneling.

As a consequence of this explanation, it was demonstrated that this effect
is just a statistical effect related to the fact of having, as a main
characteristic of the quantum formalism, probability densities with non-zero
root mean square deviations, implying that it will always be possible to
find, within such an {\it ensemble,} particles with high values of the
momentum. Even if this is highly improbable in a particular case, it is the
probability of finding such particles which will determine the effect of 
''tunneling''.

As was shown this effect may hardly continue to be called tunnel effect
since there is nothing being ''tunneled''.


\begin{thebibliography}{99}
\bibitem{1}  Olavo, L.S.F., quant-ph archieve \# 9503020

\bibitem{2}  Olavo, L.S.F., quant-ph archieve \# 9503021

\bibitem{3}  Olavo, L.S.F., quant-ph archieve \# 9503022

\bibitem{4}  Olavo, L.S.F., quant-ph archieve \# 9503024

\bibitem{5}  Olavo, L.S.F., quant-ph archieve \# 9503025

\bibitem{6}  Olavo, L.S.F., quant-ph archieve \# 9509012

\bibitem{7}  Olavo, L.S.F., quant-ph archieve \# 9509013

\bibitem{8}  Olavo, L.S.F., quant-ph archieve \# 9511028

\bibitem{9}  Olavo, L.S.F., quant-ph archieve \# 96010390

\bibitem{10}  Olavo, L.S.F., quant-ph archieve \# 9601002

\bibitem{11}  Olavo, L.S.F., quant-ph archieve \# 9607002

\bibitem{12}  Olavo, L.S.F., quant-ph archieve \# 9607003

\bibitem{13}  Eisberg, R. M., ''Fundamentals of Modern Physics'', (John
Wiley \& Sons, Inc, 1961).

\bibitem{14}  Pauling, L. and Wilson Jr., E. B., ''Introduction to Quantum
Mechanics'', (Dover Publications, Inc, 1963).

\bibitem{15}  Gasiorowicz, S., ''Quantum Physics'', (John Wiley \& Sons,
Inc, 1974).

\end{thebibliography}
\end{document}